# The Multimedia Product
# - between Design and Information,
# Design and Utility and
# Design and Entertainment

**Assist.Prof. Dieter Penteliuc-Cotoşman,**
Faculty of Arts, Dept. of Multimedia Design,
The West University of Timisoara, Romania

**REZUMAT:** Lucrarea învestighează alternativele posibile de rezolvare a problemelor legate de necesităţile de comunicare coerentă şi eficace ale oricărui produs multimedia. În esenţă, prezentarea se va axa pe identificarea aspectelor şi principiilor ce guvernează cele trei tipologii ale design-ului — în fapt, ale proiectarii multimedia, într-un sens mai larg —, şi anume : design-ul informaţiei (engl. Information Design), care vizează tocmai modurile de organizare şi de prezentare a informaţiei într-o formă semnificantă şi utilă ; design-ul de interfaţă GUI (engl. Graphical User Interface Design), a cărui sub-domeniu este constituit de organizarea informaţiilor afişate pe ecranul monitorului şi de interactivitatea dintre utilizator, calculator şi dispozitivele electronice circumscriind, de fapt, tot ceea ce utilizatorul vede,atinge, aude şi toate elementele cu care acesta interacţioneaza ; design-ul grafic, a cărui principală preocupare este aceea de a crea un aranjament compoziţional estetic (din punct de vedere vizual şi perceptiv) al informaţiei.

Today, Multimedia represents a real industry who's prodigious development is determined, first of all, by the unprecedented convergence, at a global scale, of the four *Information's Mega-Industries* — *personal computing, mass-consumption electronic houseware, printing and publishing industry, telecommunication and entertainment* —, which are, all, striving to include and use, for their products, a *common digital format*. This format is predicted and foreseen to be as universal accessible and usable, in terms of space, time or receiving / playing equipment. The beginning of this unparallel quest, for this common multimedia format, is placed in 1992, the year when the main actors of the Computing and Electronics Industries scene — giant companies such as IBM, Microsoft, Apple, Sony, Dell, Nintendo —, began to spend huge amounts of theirs financial resources in order to bring, on the specialized market, equipments and multimedia products for their own advertising purposes. Since that year, this process never ceased; on the contrary, it continues dynamically, even today.

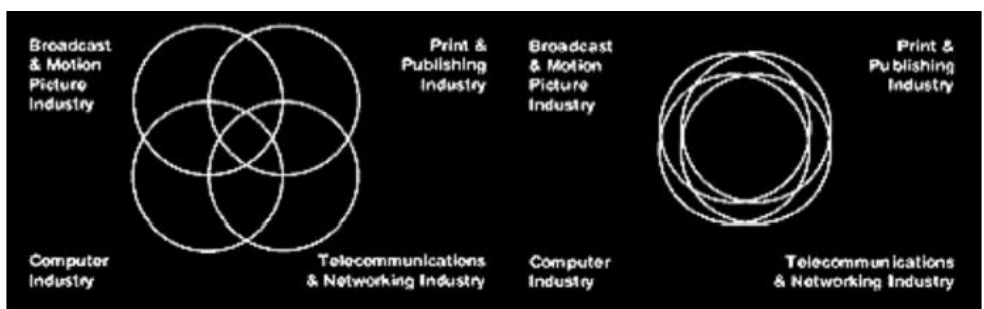

Multimedia can be defined as a continuum of applications and technologies which allow and enable a wide series of sensorial experiences; applications which are having the capability to associate and to mix different types and classes of Media: *text, illustration, photography, sound, voice, animation, video etc.,* who's content is interactive and, therefore, it actively involves the final user, yielding the control in his favour. The most important advantage that multimedia brings — answering, actually, to a more global and major tendency, the tendency to adjust and customise the content of the product to the personal needs of the user —, is the fact that it allow providing the user with a customised content, a content that is "shaped" on the "calibre" of every individual user/consumer, a content which is controlled and manageable interactively. The user can choose or select the information which is relevant for him, and for him only and, within the confines of this process, multimedia facilitates the interaction with the digital world, in order to make the sensorial received information to be as significant and charged with meaning as possible.

Also, multimedia constitutes a solution for amplifying and improving the way people are working, learning, playing and, most important, how they communicate with each other. Imposed by the Internet, the multimedia product or, better said, the multimedia artefact, represents the newest communication standard. His obvious advantages are many — the user is given with a high level of control, the developer is facing endless possibilities to present the information under various types of stimulus and digital formats and editing technologies, these many possibilities are making it easy for the user to recall and/or to memorize their content, saving time, resources and a great capacity to captivate and to stimulate the user —, and all of them serves as justifications for using this kind of products in various fields and contexts.

## 1. The Graphical User Interface — central issue within the process of multimedia design

Essentially viewed, the design of a multimedia product raises three main problems: designing the interface, the interactivity and designing the searching/displaying search results mechanism. Among these problems, designing the interface — which can be described as being the presentation mode and the behaviour of a computer system —, constitutes one of the system's face with which the user is in constant contact. Actually, the interface represents the only communication mode between users and any computer system or computer product, aspect that justifies the reasons for which the design of the interface is so carefully carried on, during the process of designing multimedia artefacts.

The Human-Computer Interaction can be eased or, on the contrary, can become more difficult, by the Graphical User Interface (GUI). The GUI should avoid disorder and the overcharging layouts that, both, lead to confusion, it should comprise only the essential elements, the compulsory needed ones and it should be appealing and attractive. Those operations done by the user frequently should be simple, in terms of interface interactions, and those done rarely and dangerous should be complicated, in order to protect the system and the users experience. GUI should not hide anything from the user and it should be designed with the principle "what you see is what you get" in mind. GUI must, also, assure the user an active involvement and participation in all programme's actions, actions which are started and controlled by the user and not by the machine, to offer the user enough feedback, expressed in a manner that is understandable by the human operator, in clear and concise terms/figures, about the progress and development of every operation. In nowadays, the majority of computing systems are conceived for being used by users with basic abilities in operating computers and all personal computers have output and display devices which are able to display colour high-resolution interfacing screens and support direct screen mouse-location and/or key-based interaction. Under these circumstances, the old alphanumerical interfaces, based on languages, much more difficult for computer non-initiated, were replaced by GUIs. Through interface graphics — colours, special fonts, 3D representations etc. —, that are giving a wider creative freedom while being designed, the creation of rich, distinct and efficient interfaces has become possible. This kind of interfaces and the various interaction techniques, developed lately, facilitates the creation of informational systems and products that are easy to learn and simply to use. The GUI interfaces are available on all personal computers, be it Windows based PC, Apple or Unix.

Their main characteristics are summarised in the following table:

*Table 1: Characteristics of a GUI interface*

| Interface's Characteristic | Description of the specific characteristic |
|---|---|
| **Display and dialogue windows** | Multiple windows allow various types of information to be displayed on the user's monitor screen. |
| **Icons** | Icons represent different types of information: on some systems, they can represent files while, on others, they can represent processes. |
| **Menus** | The functions are selected from a roll-down menu, instead of being entered from a command-line filed, in a command language. |
| **Location devices** | A location device (e.g. the mouse) is used for selecting an option that is available as displayed on a menu or is used for pointing at the elements of interest from a dialogue-window. |
| **Graphical elements** | These kind of elements may be mixed with paragraphs or/and text titles, on the same display surface, |

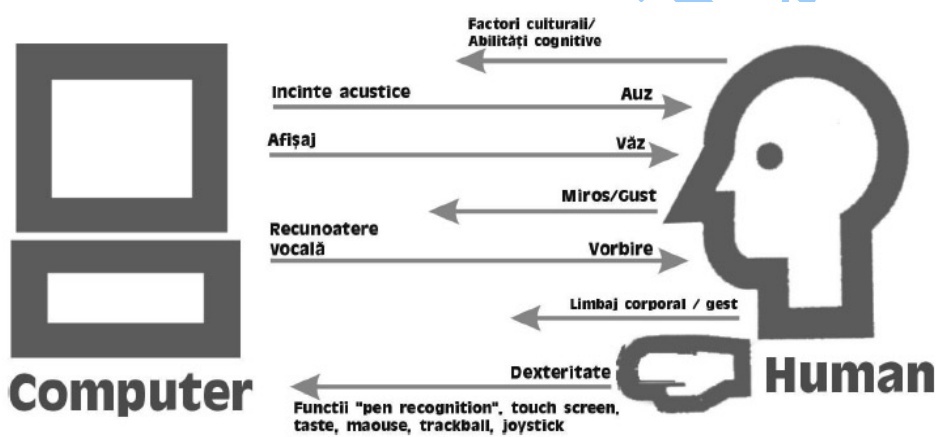

The GUI interfaces offer few remarkable advantages:
- They are, relatively, easy to be used and learned, even for PC beginners, these kind of users being able to use a GUI interface after a short introductory session;
- The user has at his/hers disposal, in order to interact with the system, more than one screens and dialogue-windows, passing from one task to another is possible without loosing, from his visual field, the information which were generated while the first task was performed;
- The user can rapidly interact, on the whole surface of the screen, having access immediately to any of the locations/functions contained by the interface' elements.

The multimedia products use only GUI interfaces and, among, it uses with predilection those based on object-oriented programming languages. The GUI interface in a multimedia product, like the interface of other systems and informatics products, has to work as a windows and menus management structure and contains a collection of commands which the user can exploit in order to interact with the product. User's communication with the multimedia artefact is accomplished in two manners: *graphically and textually*.

The language of the interface must be simple enough in order to be understood by the user, as it has to be efficient and complete, and it's grammar must be a natural one, meaning that it will be based on minimum number of rules, easy to be followed. Also, the user interface should allow undoes and returning to the previous situations, when mistakes and errors are committed by its user.

## 2. Types of Graphical User Interfaces

The most common and knew types of interfaces are those based on direct-object handling (direct manipulation) — which allow user's interaction through direct information alteration/changing that is visible on the screen —, and the menu-based interfaces — which contain particular and specific commands.

An interface for direct handling presents to the user a model of its informational space, and the user interacts with the existing entities in that particular space through direct actions, as the replacement of information or changing information's location. Due to the fact that the changes done to the given model will produce chain-linked alterations on the adjacent informational layers, explicit commands are not necessary for modifying the information. The majority of text processors or the screen editors give a very familiar example of a direct object manipulation. One of the most simple and easy to understand pattern of an interface with direct-manipulation is a form type interface. Another example may be considered the graphical interface in whose space the user is presented with a list of relevantly named files: in order to change the name of a file, the user will select the displayed text and will enter the new text, replacing the initial file's name.

The direct-handling interfaces present three important advantages: the learning curve of how to usage of this type of interface is relatively small; the users are having the feeling of being in control over the computing system and they are not at all intimidated by the complexity which governs it; the users are receiving immediate feedback after the actions they undertook, together with the errors which can be detected and corrected, almost in the same rhythm. The direct-handling interfaces are complex interfaces, from the programming and hardware requirements (RAM and processor capacity) but, after the huge break trough done during the '90s, these requirements stopped being a problem.

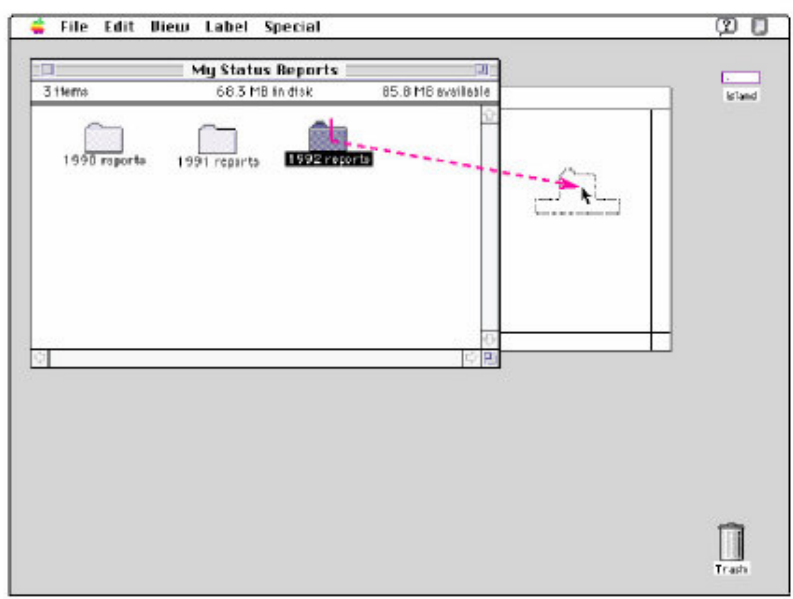

In the case of a menu-based interface, the user is selecting one from the multiple possibilities for launching o command to the computer. The user may enter the name or the selection's identifier, or can point with the mouse (or another locating/pointing device), he can use the cursor movement arrows in order to place it on top of the selection and, on some touch-screen terminals, and the user can point the selection with the finger or a stylus-pen. Menus are of two kinds: pull-

172

down and pop-up. The first type o menus display the title of the menu whose selection pulls-down all the comprised commands in order to select one of them. The pop-up menus are associated with different entities — a field or a shape —; the selection and a click of that mouse would make the whole menu to become visible. The pull-down menus bring the advantage that the user is permanently aware of their presence and he/she can anticipate, at any time, the result of their activation through clicking them. In spite of all these advantages, the menus do occupy space from screen's area, a fact that can become problematic if those functions are seldom used. The pop-up menus are adjustable: their options can be changed in order to suit better with the entity to which they are associated with. Besides this, opening a pop-up menu may contradict the principle, which states that the interface should not surprise the user who can be confused by this menu and he must invest time to understand its effects before selecting a certain command to be executed.

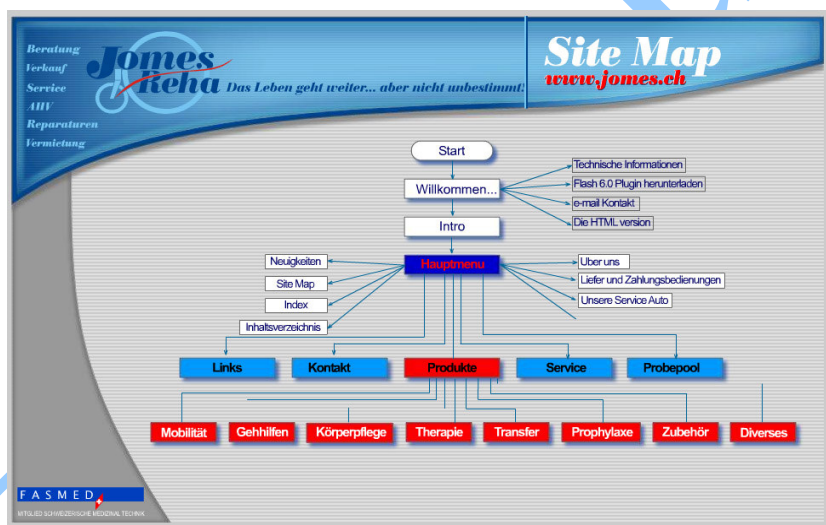

A major problem of the menu-based interfaces is the need to structure large size menus, including a huge number of options, varying between tens, hundreds or even thousands. These options require too be organized in a manner that would allow them to be displayed in reasonable small portions. In order to solve this problem, the multimedia developer can resort to the following types of menus: scrolling menus, hierarchical menus, scrolling menus and associated control panels. Navigating through the menu can be simplified by displaying a kind of menu, which is, in the end, a compressed image of its hierarchy. The map of the menu will indicate the user's current position within the structure of the hierarchy, the path the user followed before he arrived in that location and other parts of the

173

hierarchy that is to be activated or visited. This map can, also, facilitate the navigation for the user, allowing him/her to select a menu by pointing it on the map and, in this way, to make huge jumps, from one part of the hierarchy to the other. Together with these two-dimensional interfaces, the multimedia artefacts are using three-dimensional interfaces, too.

The 3D computer-aided modelling and the latest display technologies are very useful when it comes to make virtual presentations for museums and galleries where holograms and video-based items can replace the most sensitive or valuable exhibits, while enabling the user to view these exhibits a complete and dynamic way, from all the possible angles. Mixing a series of different types of media — text, photography, slides presentation, multi-screen presentation, film, sound, light —, which are all, somehow, playing a particular role in representing an idea and creating a complex, have determined the use of multimedia in various exhibition-like context. As an information source, the multimedia exhibitions have no rival and the virtual reality comes to compensate some aspects regarding a real-life exhibition, which cannot be covered by this type of virtual exhibitions. The, so-called, *stereo-books* or the *pop-up* packages (that contains 3D representations), exploded objects, dynamic diagrams, three-dimensional text, informational "landscapes" (an information presentation method that is composed from a global, general and vast view, a view made-of blocks of text placed in space, in the foreground or background) are all new means to exploit 3D computer-aided modelling techniques, much more efficient for transmitting the information than the plane 2D image will ever be. Following this line of argument, it can be said that the digital exploration of the virtual space represents, today, the latest information's design's frontier.

## 2. Multimedia Design: the *Information's Design*

The multimedia design involves three types of activities, tightly interconnected with each other: information design, interface design and graphic design.

Solving the problems regarding the communication necessities is an essential aspect for any multimedia product. The information design, which includes all informational media — textual, aural, visual —, and their interactions, is aiming at the organisational structure and presentation of the information, in a useful and significant form. Unlike Graphic Design, whose main preoccupation is to create an aesthetical arrangement of all informational elements (beautiful, in terms of visual perception), the Information Design pursues to create, first of all, a clear, exact and meaningful arrangement with the informational parts Besides these differences, this does not place the two disciplines in opposed positions. On the contrary: it would be ideal an information which is attractive and meaningful, in the same time. Information Design may resort to colour alterations, composition and styles of any medium changes, but it's main concern remain the organization of all elements of a product, as a whole, as a unity, because organisation affects the communicating capacity of these elements. Every organisation mode of information answers to a certain necessity. Richard Saul Wurman (Information Anxiety) identifies five general modes for organizing the information:

- *Alphabetically* — the most common form for organizing the information, the most familiar and, generally, the most easy to understand, at least in the western culture, although it has no meaning in it self;
- Organisation *based on timelines* (chronologically) — it is an efficient manner to structure information, especially for establishing time-based relationships among a number of singular events and, due to it's universality, as long all the known cultures are sharing the same measuring time methods;
- Organisation *based on space location* — a common form for presenting the space relationships, very efficient when the message contains such relations among its components;
- Based on two, opposed parameters — this kind of organisation is establishing hierarchies, levels, grades;
- Based on *categories* — a type of organisation that applies a certain cropping of the information, whose content is divided and grouped, depending on the communication intention.

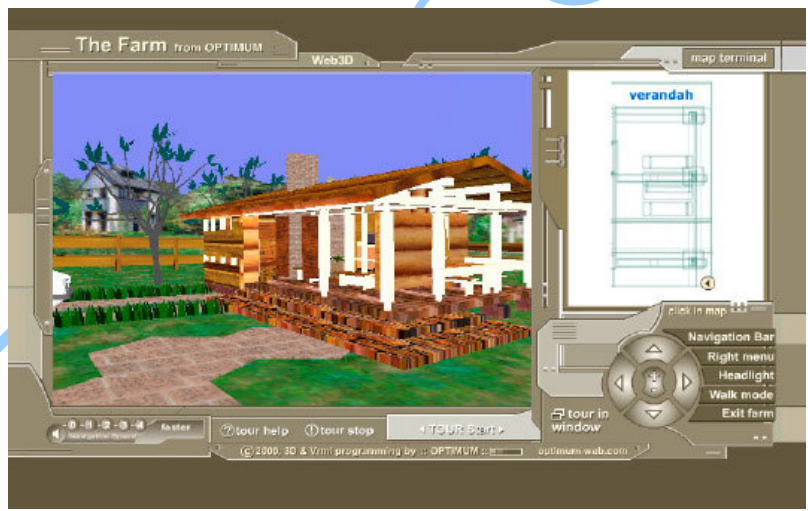

Every informational product must have only one, clear and obvious way of organisation but it may have, also, sub-organisations. Also, the organisation of it's various parts and sections may be different. Depending on the nature and the content of the product, more than one organisation manners may be combined. It may be useful to include a supplementary, alternative organisation within the main organisation principle, under the form of alphabetically ordered indexes, depending on their location, in categories etc., indexes that are offering different criteria and searching ways, answering in a better way the users' needs and objectives.

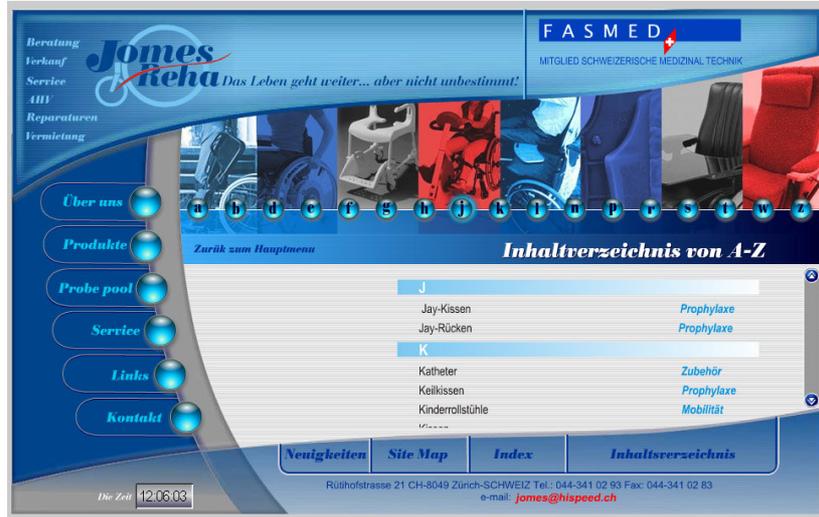

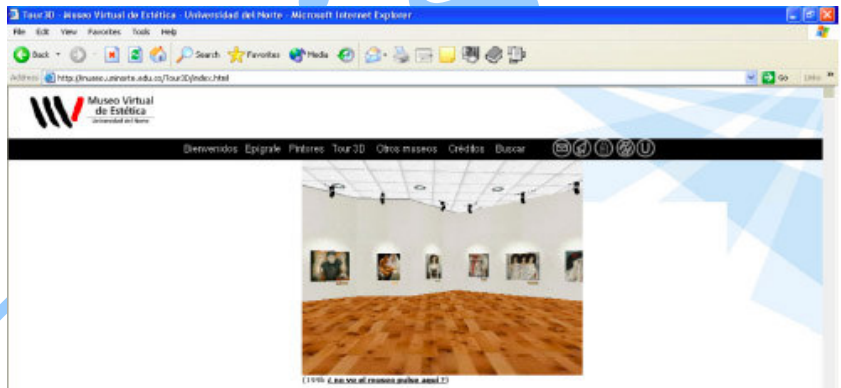

The organisation mode of a multimedia product, including it's secondary and tertiary organizations and the multiple accessing ways of the product's content, must express a certain communication intention that is present in this content and aimed by it. The Information Design resorts to few instruments, each of them adjusted to a certain function: a map or a diagram are useful to describe the interaction or the relationships between different elements; some sketch describes well the order and the succession; an informational structure presents, at an information level, the elements and the connections among them, especially when the number of connection is bigger than the linking points; in film production,

176

theatre or video, the storyboard is a useful and handy instrument for visualisation (the storyboard represents an illustrated plan, structured on a certain timeline and it is conceived on scenes which are telling a narration and which is establishing, in every scene, the weight of the visual and aural elements. The sensorial and intellectual experience given by a multimedia product is similar, in many respects, with the one created by film and theatre, because in all these situations, multiple forces are combined to generate a rich, multi-sensorial, communicational experience.

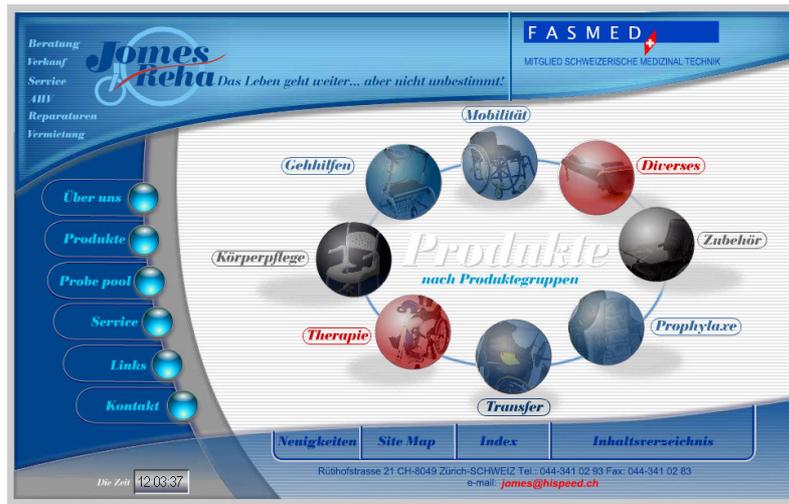

## 3. Multimedia Design: the *Interface's Design*

Related, to a certain extent, to Information Design, Interface Design has, however, it's own realm, constituted from the displayed information on the monitor's screen and the user's interactivity with computing systems and their peripheral devices. Under the sphere of Interface's Design, can be placed everything the user sees, touches, hears and all the elements with whom the user interacts with, therefore, the ensemble of audio, visual and tactile components together with the interaction and navigation. The Interface's Design orients the user for understanding the message and for updating the experience suggested by the multimedia product. Paradoxical, the ideal interface is the one that remain "transparent" toward the user, transparency in this case means that the interface is not distracting user's attention from the scope or/and the message of that specific experience. It means that the interface is so subtle that the user does not perceive it. The elements of an interface form a rich series, starting with the screen's layout and colour selection and goes until different interaction modes — keyboard keys, touch screen, joystick and vocal control.

The Interface Design is a challenge, because it supposes not only combining and balancing of the various communication modes, but also of the communication traditions, in order to realize a meaningful and understandable experience. The Interface's Design starts with the experience type one would like to achieve and takes as a guide two fundamental principles: first of all, the interface should not obstruct or restrict user's intelligence and his /hers abilities and faculties, nor to constantly reminding him that fact that he is working, in fact, with a machine; secondly, the interface should give the user as much control as possible over the medium in which the experience is taking place and to make it accessible.

The basis from which, the design's of interface, starts is the understanding, on one hand, of the user-computer interaction and, on the other hand, of the user's thinking structure and functioning, together with the way in which he/she is perceiving his/her environment. Designers must take into account the needs and capacities of users, their physical and mental limitations while using computing systems. Maybe, one of the most important limitations, among these, is about the size of long and short-term memory.

According to one of the most famous models of the human memory, described by H. Gleitman (Psychology), all humans store information in to an area of memory called short-term memory (STM) or in a system called long-term memory (LTM). The STM memory system can cope only with a small quantity of information (7 memory units + or – 2), the memorising time for this memory is short also (approximately 15 sec.) time after which, if the information is not processed in STM it will get older and forgotten (or lost/erased).

Unlike the STM, the LTM has an enormous capacity, and the information that is contained can be activated by elements like words, smells, and sounds, all-powerful enough to make the individual remember a significant quantity of information. The way the memorising system work has significant implications in working with the computer. Due to the limited capacity the STM memory, all information needed for taking decisions should be displayed, simultaneously on the screen, a situation which asks from the user to keep this knowledge in STM or to take notes, activities which will lead to useless memory overload and work slowing down. A simplified model of how information is processed by humans [Ras83] shows the fact that cognitive processes operate at different perception levels. While, at a conscious cognitive level, one single process is possible in the same time — this level is reserved for reading and semantic information understanding, complex problem solving —, at a lower level of understanding, it is possible to perform more than one operation in parallel and almost automatically, without making cognitive efforts. Regarding the way the user's interface should be designed one aspect must be stressed out: the user should be capable to use it almost automatically (at a lower cognitive level), leaving the higher cognitive levels for solving the complex problems that might arise [NAL92].

The sensation's and the perception's study put into evidence the fact that humans are perceiving the objects as well organized entities and not as separate bodies, the perception it self, being founded on recognizing models. "Model", is understood here, as a "complex composition of sensorial stimuli which an human observer is able to recognize as a member of a class of objects" [So91]. The knowledge of perception, of its organizing principles — proximity, similarity and closure —, are essential for designing an interface, about what concerns taking decisions about colours, fonts, sizes, grouped information placement on the screen, in order to optimize the reading and searching processes. Experimented users are able to decode significant models rapidly. If a collection of objects will always have the same placement on the screen, global models may be obtained for guiding the reading process. Also, decoding a model is possible, if the variable has always the same colours or/and shape.

The limited human capacities are placed at the foundation of GUI design principles and they are : consistency, minimizing the need to memorize, providing feedback, reducing the error occurrence, giving the possibility to return after an error context, several level of experience possibilities.

The consistency of an interface refers to the fact that all system's/multimedia product's commands and menus must present them selves to the user in an identical format, as it refers to the fact that all parameters can be transferred to all commands in the same way and the commands' punctuation and syntax are similar. The consistent interfaces reduce the time needed for learning of the usage mode of a system or a product. The interfaces' consistence is needed also at sub-system levels : as much as possible, commands with similar meaning but placed in different sub-systems, must be expressed in the same way. There are levels and degrees of consistency and the total consistency is neither possible nor desirable.

The Interface's consistence can be achieved by defining a coherent model for the interaction between the user and machine, an analogue model who belong, somehow, to the real world and the user is able to understand it. A consistent interface will integrate the system-user interaction but also the presentation of the information originating from the system, through the same generic framework, of the very same model. In order to define and name this model, it has been developed the Interface Metaphor Concept.

One of the most known, and used, metaphor is the «desktop» metaphor, a metaphor through which the user's screen is assimilated with the image of this space, taken from the real world. Destroying and entity is similar with it's moving to a trashcan (Mac OS) or a recycle bin (Windows OS), reading the electronic messages is fulfilled by "opening" a postal inbox, the files are archived in a folder and these folders are classified. For the first time, the desktop metaphor was implemented on Xerox Co. computers and, under the form of some variations, was the ground idea for the Operating Systems designed by Apple (MacOS), IBM and Microsoft.

For the complex interactive systems, the desktop metaphor is amplified through adding the «control panel» metaphor, which is graphically representing the system's commands. This control panel may take many shapes and sizes, from a group of icons, each icon for one particular command, to complex control panels, which are reproducing hardware systems control panels which can include additional objects that are particular for user interfaces : buttons, switches, display fields, indicators, sliders etc.

Using metaphors, based on the experience earned from the real world, facilitates user's understanding and experimenting of a certain type of interaction by association with another kind of interaction that is familiar. The metaphor must be clear and the whole visual, auditive and behavioural environment must be carefully built, as a stable and coherent world, for supporting the initial metaphor. Organizing an interface, based on a metaphor, can clarify the interaction but the structure of the interaction around a metaphor can be useful only if the metaphor is familiar, stable and consistent.

Efficient applications are characterized by a consistent design. Consistent in them selves and with each other, they give the user the feeling of safety and encouragement to explore. The Interface's Design, which should take into account user's need for stability, the fact that people prefer the computerized environments which stay comprehensible and familiar for those that are randomly changing, may resort to consistent graphical elements, in order to offer visual stability, and to a clear and finite set of objects and actions for operating with, in order to offer conceptual stability. Depending on how consistent the GUI interface is, it might lead to an error committing probability reduction while performing the requested tasks. However, users do make, inevitably, mistakes, when they are using a system/a product and the interface's design has the duty to minimize these errors, even if it cannot eliminate completely wrong usage and faulty operations. The interface should contain facilities that would allow the users to recover the stage before the usage error. Recovering may be accomplished by confirmation for the destruction actions (the user is requested to confirm any potential destructive action that he specifies, before entering the information), or by using an annulment facility for actions with a destructive character (the UNDO function, with one or more levels). The Interface's design should be guided by the principle of tolerance towards user's errors. Therefore, all actions foreseen by the programme should be reversible while the interface should allow the user to make any rational action, to explore the whole programme/product and to inform him/her about how to proceed for not crushing anything.

GUI interfaces should always provide the user with an on-line help (assistance). These systems represents that part of the interface which refers to the ways of guiding him/her and covers three main interest areas : messages produced by the system in response to user's actions, the on-line help system and usage documentation delivered together with the system or the multimedia product.

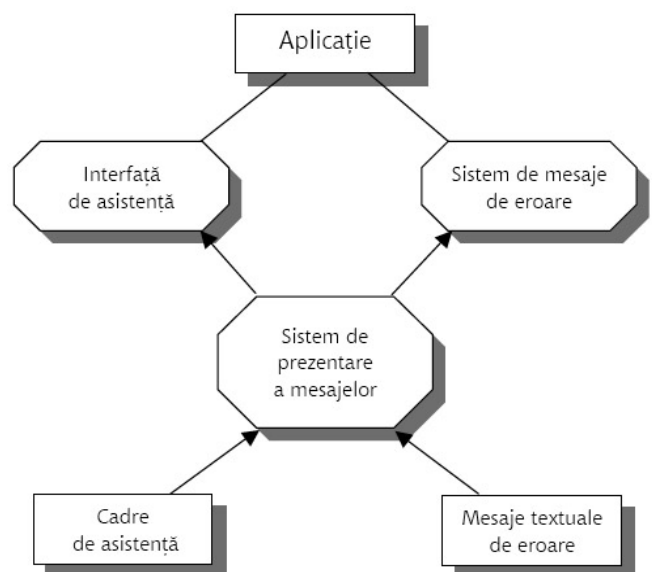

The guiding and helping facilities must be integrated with the system and at least few levels of assistance must be provided. These facilities should vary in form and complexity, from simple, system usage introductory information to a complete and detail description of systems facilities, and it should be structured in a way that the user not to be overwhelmed with information, in the occasions when he//she is just looking for help. Writing the help and on-line assistance texts or the error messages must keep take into account, on one hand, the current context that the user is finding in, his/hers experience and their level of aptitudes and, on another hand, of some style and potential users culture requirements and needs.

This assistance system must offer the chance to have different accessing gates towards it self, which should let the user, access it, at the messages' top of its hierarchy, while searching for information. Alternatively, the user can access the on-line system for receiving an explanation or an error message or it can demand for an explanation regarding a certain command included with the application/product. All well developed on-line help system do have a network like, complex and hierarchical structure, within which every frame of assisting information is connected to several such help frames.

The *hypertext* and *hypermedia* systems can be easily used for implementing help message-based systems. The *hypermedia* systems within which the text- or image-, sound or video- based information is structured more likely hierarchically not linear, a hierarchy that can be easily crossed by partially selecting from the displayed text. The advantage this system shows is that the help/assistance information can be provided in various ways, but sometimes it is difficult to integrate it with other applications. Due to the fact that hypertext systems are context-sensitive, an aspect which asks them to be controlled by the application's contextual information.

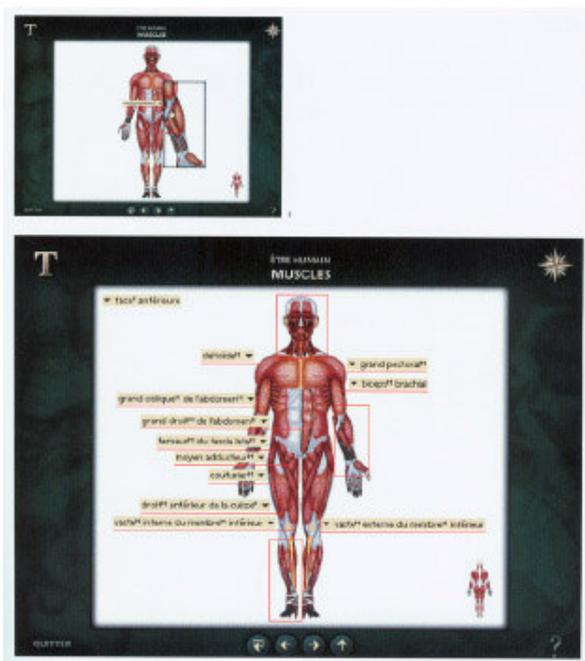

The interface design must take into account not only user's cognitive and perceptive capacities and particularities, but also his/hers expectations, needs and desires like: the need to feel as having the control over the computer's activities, the desire to get physical actions in response to user's physical actions, to get feedback from the instruments that the users works with, the minimum effort tendency, need for stability and finding some familiar elements, on which the user can relay.

The Interface Design must take into consideration, also, other aspects like *perspective, point of view, agents and on-line guides*.

Perspective (or the Voice), which might be the 1st, 2nd or 3rd, do influence user's experience : the perspective at 1st type voice gives the user an individual and personal experience, allowing him/her too make selections in order to navigate within that particular multimedia product ; the perspective at 2nd voice will engage the convert the user in a participant at the presented experience; the perspective at 3rd voice creates the sensation of detachment and objectivity, therefore seems to be more credible and corporate official like.

The *point of view* gives life to the presented content, putting it into context and giving it character and personality, but involves a certain degree of bias, favour and prejudice, and it may influence the communicated information or the way the user understands it. For this particular reason, it is good that any point of view to be counterbalanced with the opposed point of view which will enrich its significance and interpretation. In the case of multimedia products for instruction and education, the multiple points of view are an important characteristic that comes to stiffen the idea of multiple information's presentation and interpretation ways and, because of this, giving the user the opportunity to learn selectively from different sources. The «agents», involved in the interface's design, are a function with unique resources, with special knowledge, acting on behalf of the user, replacing him/her, guiding the user through interaction, interpreting the conditions requested by the machine, providing information or/and indications. Agents, appearing as «helping hands» or aids, present the advantage that they simplify some processes for the user, sparing him/her from the obligation to fulfil some repetitive tasks which involves a lot of work, as filling a questionnaire designed for database interrogation or searching in large mass of data. Agents have structured, yet modifiable, structure, which acts based on information or executing specific functions in the background of the project and they can fulfil tasks with very little input from the user who does not need to understand or learn it's sometimes complex functioning modes. Agents personification, that can be anthropomorphic or not, might facilitate user's interaction with the project, if the chosen personality is suited with it's nature and objectives.

      Another important aspect, for designing an interface, regards *navigation*. In order to meet navigational requirements, an interface must provide clews that should allow moving forwards and backwards and exploration. The way the user navigates depends on the multimedia programme's degree of interactivity : in a simple, linear presentation, the navigation path is back and forward, user not being able to do nothing but moving between these two directions or to jump over a number of screens for returning to the initial screen/page ;  programmes with complex paths are much more difficult to navigate in and they require the use of different tactics — using colours for mark the section of the project, using icons, sounds or of a set of maps that specifies the territories —, which will help the user against getting lost. Navigation is facilitated also by the use of a certain terminology, of links and consistent searching mechanism. Evan in the case of hypermedia means, which are offering the user more than one way to interact, the interaction must always be clear and navigation must follow a certain structure.

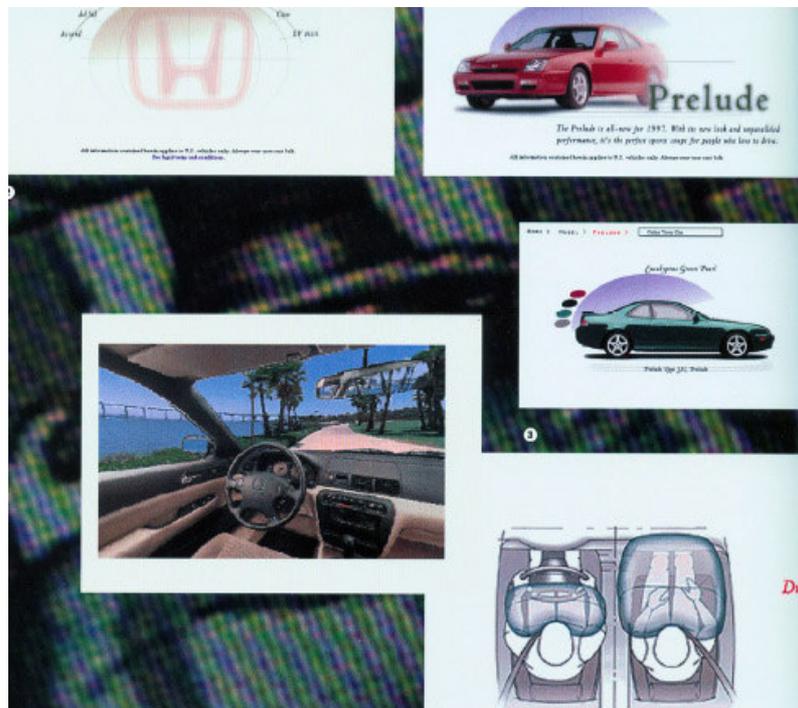

One of the basic rules regarding navigation says that the user should not made more than three "jumps" from his/hers initial request until the desired result. Another navigation "golden" rule require that the user should always know where he/she really is, within the products framework. A good navigation assumes a consistent placement of logos, frontispieces, titles describing the position and the hierarchy of a certain screen in the project's perspective viewed as a whole. Also, the user should be able to cancel any action he made and return, without difficulties, to the location from where he/she left.

Another criterion that must be taken into account, when designing a GUI, is the usability factor. Usability, which reflects to what extent the system or the multimedia product can be used in order to accomplish a certain scope (aim), comprises two categories: utility — which proves that the needed functionality is included in the system/multimedia product —, and usability — which points out to what extent the user can exploit this functionality in order to reach his/hers objectives effectively and in a satisfactory way, within a certain context of utilization. Usability refers at the following attributes : the learning's difficulty degree needed for the user to use the system productively ; operating speed ; robustness, regards the easiness with which the system can cope with user's errors ; recovery degree or the system's/or product's easiness to recover after the errors committed by the user; adaptability.

Colours are a very important issue for interface design. It gives a new dimension that can be used for displaying on the screen complex informational structures. In other cases, the colour can be used for attracting operator's attention to some important events that have been detected by the software.

Schneiderman [Shn92] offers few instructions regarding how colour should be used efficiently when designing GUIs. We made a selection of those that we consider are the most relevant and important to our line of argument:

- Reducing the number of used colours and adopting a conservative attitude about using them. It shouldn't be used more than 4-5 different colours in a single window and 7 in a system's interface.
- Using the colour in a selective and consistent manner, not only for render it agreeable. The colour has the capacity to indicate, for example, a changing in system's state. If a control display changed its colour this should mean that a significant event took place. Stressing out through colour is a very important aspect for the complex types of displays within which hundreds of distinct entities may be visualised in the same time.
- The Colour-based Coding made in order to support the task that the user is executing. For example, if the user must identify abnormal instances, it is good that these will be stressed out; if similarities are to be discovered, than they will be emphasized through different colours.
- The careful and consistent use of Colour-based Coding. If a part of the system displays red coloured error messages, all the other parts of the system must do the same way. The presuppositions that the users might have, regarding the meaning of certain colours, should be taken into account.
- Careful association of colours. Because of eye's physiology, the humans are not able to concentrate their visual sight on red and blue simultaneously and other, contrastive, colour combinations can be visually disturbing and hard to be perceived.

If it is correctly, rationally and consistently used, the colour has the capacity to improve the GUI, helping the user to understand and to perceive complexity. Many of the existing computer systems overlook the fact that the computer systems' users, are trying to resolve a certain problem while being helped by machines and these systems do not take into account his needs and limitations. The Interface Designer must always remember that the system's users do have a task to accomplish and the interface should be oriented towards reaching this objective/task. The computer system's potential users are allowed to directly participate within the interface design process, as members of the design team. This approach has been tested successfully, in Scandinavia, where it is called "participatory design" [GC91].

The Design of a GUI interface can be approached from a number of angles and perspectives. Wallace and Anderson [WA93] identified four such major possible approaches:

- from the *craftsmanship perspective*,

- from the *cognitive and technological engineering* perspective,
- from the *advanced software engineering* perspective.

According to the *craftsmanship perspective*, every project is unique, and the use of general interface design methodologies becomes impossible. In exchange, other abilities are required. This approach assumes that a good project comes from a good designer.

The Cognitive Engineering is an attempt to apply the information processing theories and problem solving to interface design. The engineers are trying to realize the programmers' discharge, from the hard and time-consuming task of interface design, by putting assisting tools for design process at their disposal. The approach from the advanced software engineer requires that the task analysis methods should be entered as an extension of software engineering methods, in order to help the design process.

Interface designing is a creative process of building and optimizing the GUI, based on the relevant computer system's user requirements and characteristics: knowledge, abilities, experience, professional preparedness and physical attributes. The rules and recommendations which describe the interface's presentation and functioning modes, varying with style, the technical medium — screen's dimensions and resolution —, the interface's integration tool, are the main aspects the interface designer should keep in mind.

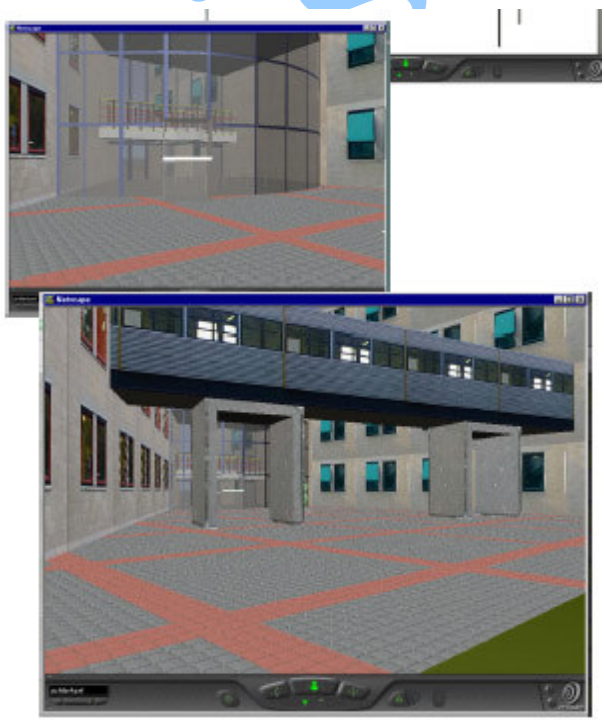

Although there are no general usable solutions and modes for designing a GUI interface, few methods that facilitate this process can be identified. One of these methods is the Design rationale Method, who's scope is to support the decision making process and the documentation of the taken decisions, within the design activities, being very helpful for structuring the relevant information. The design problems, options and the applied criteria are specified for every taken decision. A typical design problem could be to decide upon the selection procedures for the various operations inside the interface where, for every design issue, different design options are identified (e.g. selecting a function from a menu or from a button). Every option is documented, together with a list of criteria only to ensure that the optimal design option is selected.

The design process can be significantly improved through continuous realization of prototypes that are tested by the final users in an iterative (trial&error) manner. The realization of prototypes is, essentially, an empirical approach to interface design, but it allows testing the design solutions and avoiding the potential usability problems, as early as the first interface design stages.

## 4. Multimedia Design: the *Graphic Design*

The realization of a GUI interface require various knowledge from several domains, such as *software engineering* or *psychology* but, also, *artistic* capacities and aptitudes. Thus, the interface designer — who has to create a structure, in the same time, functional and visually appealing, liable to certain criteria, such as the use of standard elements, security or limitation to a certain budget — can be compared with an architect who is, half engineer, half artist.

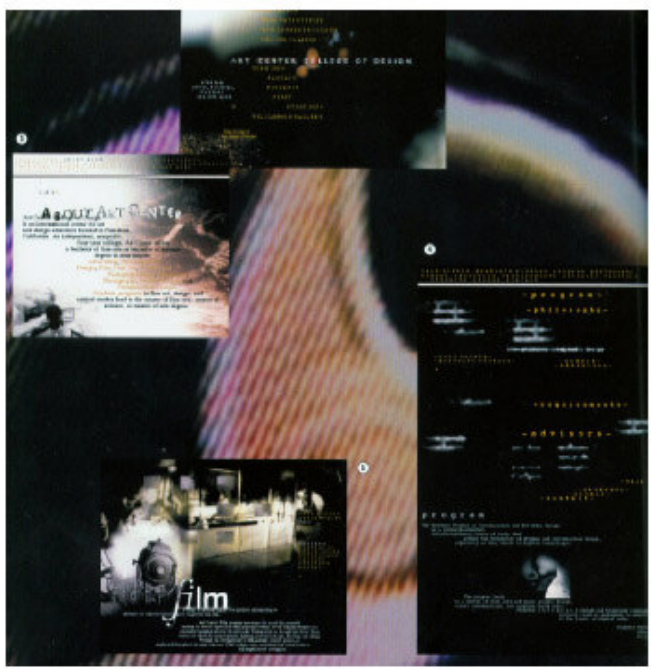

Aesthetics is, together with efficiency and usability, an important criterion for the GUI interfaces design. The multimedia product has to be, not only functional, but pleasant and aesthetic as well. It must transmit to the client, by it's

187

presentation, what kind of product it is, what is the context in which it will be utilized and what is supposed the user to expect from it. Also, the presentation of the product can be essential for the utilization comfort and often becomes a competition issue between products and rival companies.

The aesthetic of a GUI interface falls under the incidence, or the jurisdiction, of Graphic Design or Presentation Design.

The Presentation Design makes the passage from the abstract conception of the multimedia artefact to its physical realization, establishing the relationships between the concept design abstractions with a real multimedia medium. Essentially viewed, the Presentation Design, effectively realize the multimedia interactive contexts. The domain and its specific issues consist in, on one hand, conceiving the multimedia artefact global framework and, on the other side, the design and the integration of all individual media.

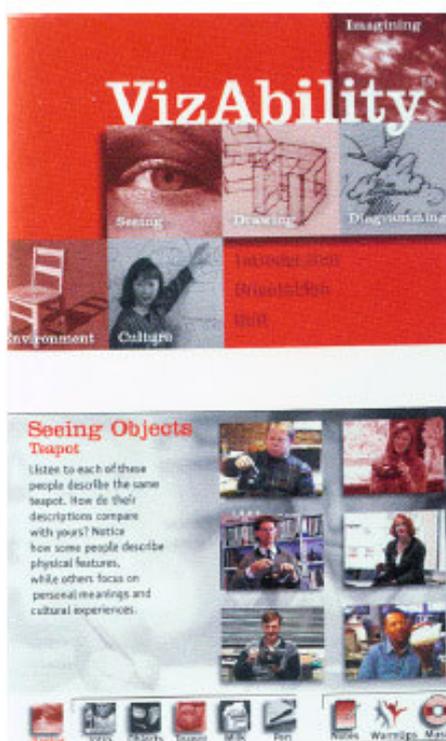

The bases of Presentation design are constituted by the main characteristics of human perception, which appears as an active process, in the same time constructive and selective. The perception involves the interaction of two information sources — the information originated from the senses and the knowledge gathered and stored in the memory — the perception process which consist of establishing relationships, in an significant manner, between the information provided by the senses and the previously gathered experience. Therefore, the perception can be synthetically defined as an active system, selective and adjustable, whose functioning is based on models.

The Presentation Design must take into account the selective nature of perception and the strong influence perception has over the expectations horizon and model recognition. One of the tasks attributed to Presentation Design is the conceive clear contexts and activities, from which the user should be able to infer models capable to determine them to shape expectations that will lead to a correct perception. Another task, which arise from the recognition of the fact that perception is strongly oriented towards models perception, would be to create interfaces with clear models, consistently utilized, that will facilitate an efficient perception of complex information.

The Presentation Design's main preoccupation is to create a coherent multimedia artefact in its totality, and one of the coherence's fundamental element, as well as aesthetics', is the composition.

The most important characteristics of the multimedia composition are unity, harmony and visual balance.

Unity is referring to the multimedia experience's totality, in the sense that all experience's components must belong to a whole, and it has to form an undivided unit. The perception of this unity derives from the system's conceptual design. It's various components are perceived as parts of a whole, because of the roles they play within the context's framework, and incorporating sub-contexts is possible because these components have clear functions in the wider context.

Harmony is referring to the way system's parts are aesthetically matching with each other and it is a concept tightly connected with the formation and the maintaining of the expectation horizon. Thus, the basic layout of the same type of components must be kept consistently in all the system's interfaces. The transition mechanisms, that are following the model's pattern at which the user is expecting — a certain model is expected because the user knows what is the context where that model is present — contributes, also, keeping the harmony. The principle of consistency is, here too, essential, consistently placing the interface's elements, after the same pattern, helping the user to navigate efficiently within the multimedia system. A conceptual framework that is clearly corroborated with a consistent interface design helps for creating and maintaining of a harmony and unity impression. In order to maintain harmony and the interface's consistence, these last ones should follow all the same basic pattern/model, that can be liable to a subsequent elaboration to respond to other, new and bigger, informational demands and for including aesthetical variations.

The visual balance is referring to the visual weight that is attributed to every component and the distribution of these objects within the interface.

The visual composition involves also a dynamic aspect — the flux — which is referring to the manner of leading the user's sight, within an interface, from one significant element to another. When an interface is displayed, the user's attention should be attracted, first of all, by the focal object and, only then, oriented towards the other objects in an order that should emphasize the relationships among the elements. The bigger the number of objects from an interface is, the harder that is to obtain this effect. The visual balance principles and the flux contribute greatly at the multimedia elements structuring.

Graphic Design corroborates the composition principles with the materials that are available, using a series of specific instruments. There is an amount of resources that contributes to the dynamic process of multimedia composition. These resources includes :
- the structures provided by the concept's design,
- the composition principles,
- specific directions suggested,
- relevant examples,

- instruments and the afferent libraries,
- experimenting and visualisation techniques.

If the composition principles orient the passage from the concept's design to presentation design, from the abstract idea to its practical realisation, the specific directions suggested are operating at a more particular level being, often, associated with the GUI systems. They promote certain general principles and global rules that are guiding the building of GUI interface's components. Libraries, for instance, are a useful source, because they offer concrete models for how the design principles can be transposed into practice and for complete integrated systems. To illustrate them, the concrete images are coming to stiffen and to render the composition principles more accessible that might be, thus, either too abstract or too fragmented. In exchange, these principles are enabling the designer to analyze the examples and to identify their qualities, this approach ensuring a solid ground for conceiving new systems.

The instruments have two characteristics that are influencing the design decisions : idea and the incorporated design physical solutions, by default, present within the instrument — navigation buttons,, scenes for developing graphic or video images etc. —, and the libraries, very rich, in terms of pre-defined components — screen templates, widgets, which are combinations of graphic entities and programme sequences that are implementing the actions contained by these entities, and clip art, pre-manufactured graphical files.

Still, Graphic Design raises problems that cannot be solved merely on applying various principles. These problems require the invention of solutions, their experimentation and improvement. In order to see if a solution is working or not, it is needed to resort to a visual prototype of the idea, therefore its experimentation through visualisation, by using an instrument for creativity and creation. Viewed from this perspective, the multimedia design process is a very dynamic one.

A central role within the multimedia design is played by the context. In their study made in 1993, Multimedia computing : case studies from MIT Project Athena, Hodges and Sassnet suggested, leaving from the film analysis, a theoretical framework for multimedia design, and they support the idea that the two main film components — direction and montage&editing —, are to be found in multimedia design too. Thus, certain parts of an interface correspond, in rough lines, to the film scene. It is all about groups of informational resources (texts, graphic images, video) and control devices used for their manipulation, configurations that are manifested as independent units, which appear and disappear together, as Windows MediaPlayer is. These groups of objects and the afferent interactivity form a distinct context for the user, and within the framework of a multimedia product analysis, the context is the equivalent for the scene cropped from a movie.

The composition of a scene must cover two functions that are operating in parallel: to transmit the information and to create an aesthetic effect. Both functions lead to the context concept. The multimedia designer's task is, thus, to select the information that will be included into a context and to conceive the aesthetic framework of that information, guiding his/hers creativity in accordance to the information and visual composition principles, derived also from film analysis. The essential issue that is making the film scene to be different from the multimedia context, is the interactivity. The film supposes a unidirectional communication and a passive spectator, unlike an interactive multimedia project that transforms the user in an active participant. Context has an interactive aspect, because it is a mode to organise the information and the afferent interactivity. The design for interactive multimedia learning environments means (IMLE), in its essence, conceiving interactive contexts for learning.

Hodges and Sassnet have emphasized three macro-functions that are orienting the options in multimedia design and which ascertain the building of interactive multimedia learning environments:
- the content's structure,
- the interactive function,
- the composition function.

This last one, which supposes parallel options with those determined by the other two functions, has the role to ensure unity, the global character and coherence of the multimedia product.

The combination of contexts — the way that these contexts are dividing the screen's office or the manner in which the transition from one to another is made — is associated by Hodges and Sassnet with the film montage (editing) from the film theory. In a film, a scene fills the communicational space and the scenes are related in a common temporal dimension. In multimedia design, the situation is more complex, and the range of possible links that can be established among the multimedia contexts is much more wider than the links among the film scenes. The most important aspect that these differences are emphasizing is the transition between contexts, as long the main purpose is to maintain the coherence in such a way that the user will not be confused in the end. Sometimes, the user must receive an explicit support that should enable him/her to cope with the transitions from a context to another. Maintaining coherence is not only a matter of providing explicit transition support mechanism, but it depends by the conceptual structure of the whole system, a structure that is determined by the hierarchical relationships between the built contexts.

The relationships between contexts include:
- hierarchical relationships of type "a part of",
- relationships of type "is an alternative to",
- transitions between contexts which are temporary linked,
- spatial relationships.

Essential is the choice of a main type of context for the entire systems, related to whom, the specific pointed out, on the screen, contexts, will be placed in a relationship of type "part of", or they will be alternatives of a standard prototype, a fact that will offer flexibility. The simultaneous appearance, on the screen, of more than just one , must seem logical for the user. These contexts can be, for instance, functionally linked, as components within the framework of a global context.

Being aware of the fact that the issues we tackled, in the previous paragraphs, cannot be exhausted in the little space we had at our disposal, we hope that, in spite of these confines, we expect and hope that we succeed, through the content of our summarized article, to give the reader a more detailed image of the options that are chosen, by the multimedia developers, in order to satisfy the need for coherence and efficient communication of every multimedia product.